\documentclass[useAMS,usenatbib]{mn2e}
\usepackage{latexsym,graphics,epsfig}
\def \be{\begin{equation}}
\def \ee{\end{equation}}
\def \bea{\begin{eqnarray}}
\def \eea{\end{eqnarray}}
\def\etal{{et al.\ }}

\title[Dynamical friction of radio galaxies in galaxy clusters]
{Dynamical friction of radio galaxies in galaxy clusters}

\author[Biman B. Nath]{Biman B. Nath\\
Raman Research Institute, Sadashivanagar, Bangalore 560080, India}

\begin{document}
\maketitle

\begin{abstract}
The distribution of luminous radio galaxies in galaxy
clusters has been observed to be
 concentrated in the inner region. We consider the role of 
dynamical friction of massive galaxies ($M\sim 10^{12.5}$ M$_{\odot}$), 
assuming them to be hosts of
luminous radio galaxies, and show that beginning with a Navarro-Frenk-White
density profile of a cluster of mass $M_{cl}\sim 10^{15}$ M$_{\odot}$
of concentration $c\sim 5$ and collapsing at $z\sim 1$, 
the density profile of radio
galaxies evolve to a profile of concentration $c \sim 25$, as observed, 
in a time scale of $t\sim 3\hbox{--}5$ Gyr.
\end{abstract}
 

\section{Introduction}
Recent observations have shown that powerful radio sources in galaxy clusters
are concentrated toward the cluster center. In a survey of 30 clusters of 
galaxies,
  Morrison \& Owen (2003) found that the spatial distribution of
 high luminosity radio galaxies in rich clusters is described by
a small core radius. In particular, they found that 
  the spatial distribution of galaxies depend on their radio power---
high luminosity radio galaxies have a smaller core radius ($\sim 0.12\pm 0.02$
Mpc) than low luminosity ones ($\sim 0.4\pm 0.08$ Mpc). Recently, Lin \&
Mohr (2007) have collected data for 188 rich clusters, and they have analyzed
the spatial distribution of radio galaxies contained within them.
They found that the spatial distribution of powerful radio sources could
be fit by the `universal' Navarro-Frenk-White (NFW) profile (Navarro, 
Frenk, White 1997) with a larger concentration parameter than needed
for fitting the total mass distribution of the cluster. And similar to the
findings to Morrison \& Owen (2003), they discovered that the concentration
of galaxies in the spatial distribution increased with their
radio power.

Many explanations can be put forward for such a steep spatial distribution of
radio galaxies in clusters. It is possible that the increased number density
of galaxies in the inner region facilitates mergers and interactions that
can trigger active galactic phenomena and 
increase the radio luminosity of galaxies. Also, 
radio galaxies are mostly associated with giant ellipticals which generally
 inhabit the central regions of rich clusters. Moreover, the increased 
pressure of the hot X-ray gas of rich clusters in the inner region may
enhance the radio luminosity of FRII radio galaxies by confining the
lobes with high pressure. 

There is also a dynamical aspect of the problem besides these factors.
If the hosts of radio loud objects in general are massive galaxies, 
then  dynamical friction will lead to mass segregation in galaxy clusters,
and radio galaxies will settle toward the cluster center over many orbits.
The situation is analogous to the phenomenon of large
density of millisecond pulsars in the central regions of globular clusters.
This concentration 
is believed to be due to favorable conditions in the cluster core for
accretion spin-up of neutron stars leading to the phenomenon of millisecond
pulsars. However, it is first
 the process of mass segregation in globular clusters
owing to dynamical friction that makes massive stars settle toward the
center, thereafter
leading to an increased number density of neutron
stars in the inner regions (Meylan \& Heggie 1997).

Mass segregation by the process of dynamical friction 
is also expected to occur in galaxy clusters 
 and it will lead to a preponderance of massive galaxies toward the
cluster center. This process is likely to occur simultaneously with other
processes of galactic evolution that may make galaxies in the inner region
prone to becoming a radio loud source. To study the dynamical aspect
of the evolution of spatial distribution of radio sources, one would need
to isolate it from other evolutionary process and determine if it is
a significant process by itself. We wish to address this issue in this paper.

The main motivation to isolate the dynamical aspect of the problem 
comes from the fact that the hosts of radio galaxies are often
 massive galaxies,
irrespective of their membership in clusters or groups of galaxies.
Recent statistical studies of radio loud objects show that their hosts
are predominantly massive objects (containing black holes of mass
$\sim 10^{8.5}$ M$_{\odot}$) (Best \etal 2005). 
It is therefore reasonable
to expect that dynamical evolution of massive galaxies in clusters 
will have a significant effect on the spatial distribution of radio loud
sources in galaxy clusters, irrespective of other processes that contribute
toward increasing the radio output of a galaxy. 

Here we focus on the dynamical
evolution of the distribution of massive galaxies in clusters
from dynamical friction, and compare with recent observations.

\section{Hosts of radio galaxies}
Powerful radio sources (with $P_{1.4 GHz} \ge 10^{24}$ W Hz$^{-1}$) have 
been found
to be associated with giant ellipticals, the oldest and the most massive 
galaxies in the universe. Best \etal (2005) have discussed the mass
dependence of radio activity with the help of a statistical analysis of
the radio properties of galaxies from the 2dF survey. They have shown
shown that the probability of a galaxy containing
a central black hole of a certain mass to have a given (or larger)
radio luminosity
depends strongly on the black hole mass; the probability scales as $
\propto M_{BH}^{1.6
}$. The (volume weighted) distribution of the black hole mass in the sample
showed that most radio loud objects had a black hole mass in the range
$M_{BH} \sim 10^{8}\hbox{--}10^9$ M$_{\odot}$, peaking at 
$M_{BH}\sim 10^{8.5}$ M$_{\odot}$. The total mass of the host galaxy
of radio loud sources is related to the central black hole mass. 
Fine \etal (2006) found
 from the analysis of the 2dF QSO survey
that for a QSO host containing a central black hole of
mass $M_{BH} \sim 10^{8.4}$ M$_{\odot}$, the total dark matter mass of the host
galaxy is $\sim 10^{12.5}$ M$_{\odot}$.

Based on these considerations, we will 
consider galaxies of total (dark matter) 
mass of $\sim 10^{12.5}$ M$_{\odot}$, and containing 
a central black hole  of mass $M_{bh}\sim 10^{8.5}$
M$_{\odot}$, as examples of host galaxies of  radio loud objects in our
calculation below.

\section{Dynamical friction}
Consider a galaxy cluster
of total mass $M_{cl}$ and velocity dispersion $\sigma$. We assume that
the average mass distribution inside the cluster follows the Navarro-Frenk-White
(NFW) profile (Navarro, Frenk, White 1997), $\rho=\rho_s /((r/r_s) (1+
(r/r_s)))$, where $r_s=r_{vir}/c$, $c$ is the concentration parameter, and
$r_{vir}$ is the virial radius. The virial radius is fixed by the overdensity
estimated from spherical collapse model, which is
approximately $\Delta (z=0) \sim 100$ at the present epoch,
 for the standard $\Lambda$CDM cosmological model
($\Omega_m=0.3, \Omega_\Lambda=1-\Omega_m$) (see, e.g., Komatsu \& Seljak ). 
The 
characteristic density $\rho_s$ is then given by,
\be
\rho_{\rm s} = {c^3 \, \Delta(z) \, \rho_c(z)
\over 3 [\ln(1+c) - {c \over
(1+c)} ] } ,
\label{eq:rhos}
\ee
where $\rho_c(z)$ is the critical density of the universe at redshift $z$.
For a large range of length scales (barring the central and outer regions
of the cluster), the mass density $\rho \propto r^{-2}$. The mass distribution 
in these parts will be characterized by a near constant velocity dispersion
$\sigma$. 

We assume a concentration parameter $c=5$ for a rich cluster as suggested
by N-body simulations (Navarro \etal 1996).  
We also assume a Maxwellian distribution of velocities of 
objects in the clusters
with dispersion $\sigma$, and assume
the average mass of galaxies to be $m$.

Consider the motion of a massive galaxy of mass $M$ and speed $V$
 in this cluster. 
Its motion in the background potential
of numerous small galaxies of mass $m$ will be perturbed by random
interactions with these galaxies. These perturbations will amount to 
a force of dynamical friction given by,
\be
F_{drag} (r) =-{4 \pi G^2 M^2 n m \ln \Lambda \over V^2} \Bigl 
[ {\rm erf} (X)-{2 X
\over \sqrt{\pi}} \exp (-X^2) \Bigr ]  \,,
\ee
where $n$ is the number density of galaxies of mass $m$, and
$X=V/\sqrt{2} \sigma $. The dynamical friction will slowly change its
orbital parameters and the massive galaxy will slowly settle toward the
cluster center over many orbits.

Consider a near circular orbit of the massive galaxy. Then, for a large
range of length scales in the cluster, we have $V=\sqrt{2} \sigma$, and $X=1$.
The coulomb logarithm is approximately $\ln \Lambda \sim \ln {b_{max} V^2 
\over GM}$
(Binney \& Tremaine 1987, eqn 7-13b). For $M\sim 10^{12.5}$ M$_{\odot}$,
and $V\sim 1300$ km/s (see below for the choice of this value), we have
$\ln \Lambda \sim \ln {b_{max} \over 7.7 \, {\rm kpc}}$. 
If we take $b_{max}$ as the
cluster core radius which is $\sim 100$ kpc for rich clusters, then
$\ln \Lambda \sim 2.5$. In this case, the product $[ {\rm erf} (X)-{2 X
\over \sqrt{\pi}} \exp (-X^2) \Bigr ] \ln \Lambda \sim 1.1$, and we assume
it to be unity for simplicity.
Also, we will use the average density of matter  $\rho= nm$ in the above
expression.

If we consider an inspiraling orbit of a massive galaxy owing to the
dynamical friction from its interactions with average mass galaxies, then
one can estimate the time scale over which the distance of the massive galaxy
from the cluster center would significantly decrease. Since the 
angular momentum $L \sim MVr$ changes according to $dL/dt=F_{drag} \times r$,
we have 
\be
 {1 \over r} {dr \over dt} 
 \sim - {4 \pi G^2 M \rho \over V^3 } \,,
\label{eq:fr}
\ee
 since the circular
velocity $V$ changes little in the region of the cluster potential under
consideration (where $\rho \propto r^{-2}$ and the circular velocity 
$V=\sqrt{2} \sigma$). Ostriker \& Turner (1979) also used this equation
to study dynamical friction in galaxy clusters (see also Nusser \& Sheth 1999).

 We can estimate the time scale of dynamical friction as,
$t_{dyn}\equiv r /{dr \over dt} \sim {V^3 \over 4 \pi G^2 M \rho}$.
For $V\sim 10^3$ km/s, $M\sim 10^{12.5}$ M$_{\odot}$, and using the
matter density at the characteristic radius $r_s$ (using eqn \ref{eq:rhos}), 
$\rho_s 
\sim 10^{-26.6}$ g cm$^{-3}$, for $c=5$ and $z=0$, we have,
\bea
t_{dyn} \sim &&2.4 \, {\rm Gyr} \, \Bigl ({V \over 10^3 \, {\rm km/s}} \Bigr )^3 
\, \nonumber\\ && \times
\Bigl ({M \over 10^{12.5} \, M_{\odot}} \Bigr )^{-1} \,
\Bigl ({\rho \over 10^{-26.6} \, {\rm g/cc} } \Bigr )^{-1} \,.
\eea
It is believed that rich clusters formed at 
$z\sim 1$, judging
from the lack of evolution in X-ray luminosity function of clusters
up to $z \sim 1$ (see, e.g., Rosati \etal 2002). The
 mass distribution would have settled into a NFW  profile
by that epoch. If such a cluster contained a few massive galaxies far from
its center,  then the above estimate shows that these massive galaxies
would have spiralled toward the cluster center in a few Gyr time scale.

The time evolution of the orbit of a massive galaxy embedded in a 
galaxy cluster can be obtained from numerically
 solving eqn \ref{eq:fr}. A few representative 
cases are shown in Figure 1 in for $M=10^{12.5}$ M$_{\odot}$, $V=10^3$ km
s$^{-1}$. We compute the density distribution of the cluster halo 
for $M_{cl}=10^{15}$ M$_{\odot}$ collapsing at $z=1$. We emphasize that
the results shown in Figure 1 are 
approximate and are based on several assumptions. Firstly,
we have assumed circular orbit. But the results shown in Figure 1 can be 
interpreted as the evolution of the radial distance averaged over an orbital
period. Also, we have neglected the effect of the massive galaxy being an 
extended object.  
In reality, the galaxy will be tidally stripped, decreasing its
mass, as it spirals 
inward (see Nusser and Sheth (1999) for a discussion on this effect).

\begin{figure}
\centerline{
\epsfxsize=0.5\textwidth
\epsfbox{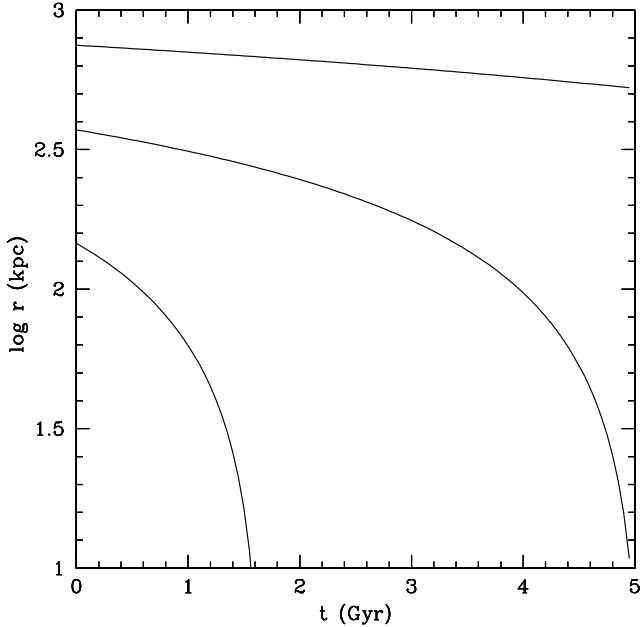}
}
{\vskip-3mm}
\caption{
The evolution of orbit for a massive galaxy ($M=3 \times 10^{12}$) 
embedded in a rich cluster  ($M_{cl}=10^{15}$ M$_{\odot}$) is shown, for
a few initial radii.
}
\label{f:evol}
\end{figure}

We can then ask how the distribution of such massive
galaxies would evolve in a statistical sense, beginning with the NFW profile
of a given concentration parameter. 
A given cluster will contain only a handful of massive galaxies, but we
can consider a statistical ensemble of clusters with the same concentration
parameter and total mass, which will contain several massive galaxies
with initial orbital parameters such that the total mass distribution of the
ensemble cluster resembles a NFW profile of the assumed concentration
parameter (here $c=5$).
Slowly, over time and many orbits, the massive galaxies will settle toward
the center, and their mass distribution will change.

To estimate the gradual change in the
mass distribution of massive galaxies immersed in a gravitational
potential determined by several average mass galaxies, we divide the
cluster into several annuli, and compute the cluster halo
density in the $i$-th annulus
as that given by the NFW profile for the average radius in that annulus. 
Then we ascribe an initial number of 
massive galaxies in each annulus, $N_M (i,t=0)$, so that the density
profile of these galaxies ($N_M(i,t=0)/vol(i)$, where $vol(i)$ is the volume of 
the $i$-th annulus) follows
the NFW profile with same concentration ($c=5$ here). We  normalize the number
of massive galaxies in different annuli
by the density at the outermost annulus, which does not change owing to 
negligible dynamical friction at that radius.

We first calculate the
change in radius of these massive galaxies initially stationed at each annulus
using eqn (\ref{eq:fr}). We compute their radii after an
elapsed time interval, and then
we re-calculate the number of massive galaxies in all 
annuli $N_M(i,t)$ at this epoch, 
again normalized by the number of massive galaxies in the outermost 
annulus.  From this we calculate the 
number density of massive galaxies in  each annulus as a function of time.
 The result
will depend on the mass assumed for the massive galaxies and the
time elapsed. Our previous estimates show that, we can expect a significant
evolution in the distribution of massive galaxies over a time scale
of a few Gyr for galaxies with $M\sim  10^{12.5}$
M$_{\odot}$.

We consider a cluster of mass $M_{cl}=10^{15}$ M$_{\odot}$
and concentration parameter $c=5$. The virial radius for a cluster collapsing
at $z\sim 1$ is calculated to be $r_{vir}\sim 2.6$ Mpc. The appropriate
 velocity
dispersion for such a cluster is taken from the scaling observed in simulations
by Evrard \etal (2008) to be $\sigma \sim 9.5 \times 10^8$ cm s$^{-1}$ 
(for 
Hubble constant
$H=70$ km/s/Mpc); we assume it to be independent of location inside the cluster,
 We consider massive galaxies with
$M=10^{12.5}$ M$_{\odot}$ and $V=\sqrt{2}\sigma \sim 1.3 \times 10^8$
cm s$^{-1}$.

\section{Results}
We show in Figure 2 the initial mass distribution by dotted line; it is
a NFW mass profile with $c=5$, for $M_{cl}=10^{15}$ M$_{\odot}$ collapsing 
at $z=1$. We plot the radius as a fraction of $r_{200}$.
Then, the mass distribution of an ensemble of galaxies with $M=3 \times 10^{12}$
 M$_{\odot}$ is shown after $t=1.5, 3$ and $5$ Gyr with solid lines. 
It is found that the distribution steepens with time.
The distribution
at outer radii does change much, but massive galaxies in the middle
region settles toward the center, increasing the density of these galaxies
in the bins
of radii in these regions. The dashed line shows a NFW profile  with
$c=25$. All profiles have been drawn keeping the mass density at the outermost
bin fixed in time, for  comparison.

\begin{figure}
\centerline{
\epsfxsize=0.5\textwidth
\epsfbox{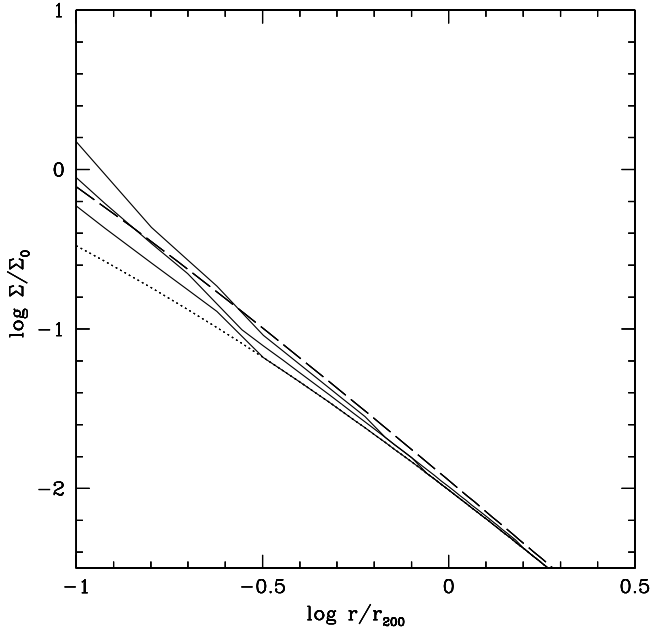}
}
{\vskip-3mm}
\caption{
The density profile of galaxies is plotted here. The dotted line shows
the initial profile, and the solid lines correspond to the density profile
of massive galaxies with $M=3 \times 10^{12}$ M$_{\odot}$ after $t=1.5, 3$ and
$5$
Gyr, the steepest profile being for the longest time interval. 
The dashed line corresponds to a NFW profile with $c=25$ (normalized
to have the same density as other profiles at the outermost bin).
}
\label{f:evol}
\end{figure}

We therefore find that the distribution of massive galaxies in a statistical
ensemble of clusters steepens with time. The steepening depends on the
assumed mass of the massive galaxies and the time elapsed. In our fiducial
case of $M=3 \times 10^{12}$
 M$_{\odot}$ immersed in a cluster of total mass $M_{cl}+10^{15}$ M$_{\odot}$
collapsing at $z=1$, we find that a distribution with initial $c=5$ has changed
to a distribution with $c=25$ over a time scale $t \sim 3\hbox{--}5$ Gyr.

It is interesting to compare this time scale with the cosmological
lookback time. The look back time to redshift $z\sim 0.5$ is $t \sim 4$ Gyr.
If rich clusters formed at
these redshifts, then
 their mass distribution would have settled into the NFW profile
by that epoch. If such a cluster contained a few massive galaxies far from
its center,  then it is  conceivable that these massive galaxies
would have spiralled toward the cluster center by the present epoch.

\section{Discussions}
Lin \& Mohr (2007) have stacked the data from 188 clusters containing
16,646 radio loud objects (with radio luminosity $P > 10^{23}$ W Hz$^{-1}$)
to produce the density profile of these objects. The number of radio
loud objects within the radius $r_{200}$ in their study
was 836. They found that the
density profile corresponds to a NFW profile with $c\sim 25\pm 7$, 
excluding the
brightest cluster galaxy (BCG). With the inclusion of BCGs which they
defined as radio loud objects within $0.1 r_{200}$, the profile
steepens to $c\sim 52^{+25}_{-14}$.

The solid curve in Figure 2 steepens further below a radius $0.1 r_{200}$,
and excluding this part (equivalent to excluding the statistics of
BCGs), we find that the density distribution is comparable
to a NFW profile with $c\sim 25$, close to the observed value. It then appears
the distribution of radio loud
objects in the ensemble cluster (produced by stacking the data
of numerous clusters) can be explained by assuming that the original
mass distribution was a standard NFW profile with $c\sim 5$, and that
massive galaxies slowly sink toward the cluster center owing to dynamical
friction. The time taken to match the observed distribution is
$t \sim 3\hbox{--}5$ Gyr if the massive galaxies have $M\sim 10^{12.5}$ 
M$_{\odot}$, and belonging to a cluster of $M_{cl}=10^{15}$ M$_{\odot}$.

It is interesting to note that Lin \& Mohr (2007) have found that more
powerful radio sources are more centrally concentrated than the weaker
ones. For example, they found that radio sources with luminosity 
larger than $10^{24.5}$ W Hz$^{-1}$ have a distribution comparable to a NFW
profile with $c\sim 59 \pm 11$. This trend is expected in the above
scenario if the probability
of a galaxy to have a given radio luminosity increases with its total mass,
which indeed seems to be the case. Best \etal (2005) found from the 2dF
survey data that the probability of a galaxy with a black hole mass $M_{BH}$ 
to have radio power larger than a given value
approximately scales  as $M_{BH}^{1.6}$. If the central black hole mass scales
linearly
with the dynamical mass $M_{dyn}$ of the host galaxy as
(e.g., Hopkins \etal 2007), then one expects the hosts
of more powerful radio sources
to be more massive galaxies.

We note here that
dynamical friction cannot be the complete story behind the concentration
of radio loud objects in clusters. Masses of galaxies alone do not determine
their radio properties; there are many factors that contribute to its
radio luminosity, like mergers and interactions with other galaxies, and 
properties of the medium confining radio lobes. Here
we have isolated the dynamical aspect of the issue, 
 and the results show that dynamical friction can also be
as important as other factors in producing a steep density profile of
radio sources in clusters.

\section{Summary}
We study the dynamical friction of massive galaxies--- which can often be
hosts of radio loud objects--- in galaxy clusters, and find the the time
scale of mass segregation to be $t \sim 3\hbox{--}5$ Gyr, for galaxies
of mass $\sim 10^{12.5}$ M$_{\odot}$ within clusters of mass $M_{cl}\sim 10^{15}
$ M$_{\odot}$. We show that this effect can help explain the observed
distribution of radio loud objects in rich clusters of galaxies.

\bigskip
Acknowledgement-- I thank M. Begelman, J. Silk and S. Sridhar for comments on 
an earlier version of the manuscript, and an anonymous referee for useful comments.

\end{document}